An Index of Science and Technology Capacity

Caroline S. Wagner, Edwin Horlings, and Arindam Dutta

Amsterdam School of Communications Research, University of Amsterdam, Kloveniersburgwal 48, 1012 DX, Amsterdam, Netherlands;

RAND Europe, Newtonweg 1, 2333 CP Leiden, Netherlands;

RAND, 1700 Main Street, PO Box 2138, Santa Monica, California 90407

Astract

Biographical notes:

# Introduction

The ability of a nation to participate in the global knowledge economy depends to some extent on its capacities in science and technology. In an effort to assess the capacity of different countries in science and technology, this article updates a classification scheme developed by RAND to measure science and technology capacity for 150 countries of the world. The first version of this index was published in the RAND report "Science & Technology Collaboration: Building Capacity in Developing Countries" MR-1357.0-WB. This article updates the index with more recent data and improves its accuracy with new data and analysis.

Science and technology capacity is defined for the purpose of this exercise as the ability of a country to absorb and retain scientific knowledge and to use this knowledge to conduct research and development. We use "country" or "nation" as the defining unit of analysis because this is how data is collected. However, we recognize that knowledge does not honor political borders, and that, even within specific countries, the "knowledge border" may involve regions that cross national borders. This would be the case between the United States and Canada, for example, where the S&T system is quite well integrated.

This index is designed as input to the policymaking process. A number of international institutions, including the World Bank and the United Nations, have policies that include the goal of enhancing S&T capacity or otherwise tapping capacity to encourage innovation. S&T capacity has been shown to be positively correlated to economic growth, although the extent to which the two are linked is not clear. (Solow 1970; Nelson, 1986; Mansfield 1968) Many areas of science have little connection to economic development, and many areas of economic growth do not rely on S&T. Moreover, a review of economic history shows that, in many cases, technology-led growth precedes the development of a national scientific system. (Freeman 2002) It may be that S&T is a *catalyst* for development; one that emerges after a basic threshold of economic development is crossed. It is not clear where this threshold is, although in recent decades, countries such as Mexico, Brazil, South Korea, and earlier, Japan, appear to have reached and crossed this threshold.

This capacity index uses a series of indicators to assess the extent to which countries have the infrastructure and knowledge absorptive capabilities to *use* scientifically based knowledge in order to enhance development. It does not measure the extent to which countries are advancing frontiers of S&T knowledge, it does not measure the capacity to *produce* scientific and technical progress, nor does it measure innovative capacity or industrial development. In addition, the index only measures comparative international differences and cannot be used to track the development of actual capacity of one nation through time. Some of these features are measured by other groups.[1]

The index uses indicators to measure the extent to which a country can absorb and use scientific and technological knowledge. Science is a way of knowing things. It is a widely accepted, adoptable, and transferable set of assumptions about how to understand the world in which we find ourselves. As scientific practices become standardized and transferred, the practice of science becomes a *system* of knowing things. That system takes on a dynamic that transcends and subsumes the practitioner. As a result, science is often referred to as a "universal" way of knowing because the knowledge passes without regard to the political allegiances, gender, race, or other attributes of the person seeking knowledge. The universal nature of science can be demonstrated by the gradual expansion of the scientific enterprise over the past 300 years from a few small centers in Europe exchanging papers and letters to hundreds of interacting centers around the world communicating through thousands of peer-reviewed journals.

Scientific knowledge is often codified in professional publications that are widely available, particularly now that the Worldwide Web offers access to full-text documents. Articles in journals serve as one source of knowledge for scientists and

---

[1] See for example the Centre for International Development, *The Global Competitiveness Report 2000* (Oxford University Press, 2000); National Science Board, *Science and Engineering Indicators 2002*, Volume 1, 2002; United Nations Development Programme, *Making New Technologies Work for Human Development* (New York, 2001); and UNCTAD, *Indicators of technology development* (Geneva, 2002).

other practitioners needing to understand natural phenomena. They also serve as a way to measure and understand the structure of and communications within science at the systems level. (Leydesdorff 2000)

Technology is a set of tools designed to manipulate the natural world and to extend human intentions. Like science, technology can be a system of knowledge, and it also includes the added dimension of purposeful, interacting tools. Knowledge about technology is often tacit – it has to be experienced in order to be learned. (Nelson and Winter 1982) Knowledge about the use of technology can also be proprietary, since the manipulation of the natural world often creates tradable goods. Thus, the measures of scientific knowledge and of technological knowledge will be different, although, in this index we do not try to capture the extent to which technological knowledge is embedded in tradable goods.

## Constructing a composite index

Before we present the index and the supporting data, it is important to describe what the index itself conveys. Composite or aggregate indices like this one are constructed for a wide range of purposes. One motivating factor is the needs of policymakers who most often need aggregated data, sometimes even a single all-encompassing number. Composite indices can be used for four distinct purposes:

- to examine a specific dimension of a field or policy area (e.g. health, well-being, economic growth)
- to compare geographic entities (usually countries),
- to chart developments through time,
- to compare social and other groups (e.g. distinguished by gender or income).

The S&T Capacity Index (STCI) examines the ability of countries to absorb, retain, produce, and diffuse science and technology. The index will essentially compare countries without measuring the actual extent of their capacity. Consequently, it cannot be used to chart developments through time (other than shifts in the international ranking). Regional, social and gender distinctions will not be measured.

The STCI is constructed in three steps, illustrated in the system diagram, Figure 1:

- Selecting variables based on an understanding of the process that determines the composite number.

- Combining the individual indicators into a single index by converting them to a common format, checking their correlation and consistency, weighting them into an aggregate number and testing different weighting schemes.

- Checking related but different variables to see if the outcome matches alternative evidence.

As we constructed the index, we kept in mind that transparency is very important to the usefulness of the index. The data and analysis presented here is done in a way that allows other researchers to apply the same index to other situations and to revise the index with new information and with the addition (or omission) of variables.

## Selection of S&T indicators

We set out to create an international comparison of science and technology capacity. However, S&T capacity is a theoretical construct. Its magnitude cannot be determined directly, let alone precisely, and has to be approached using a number of proxy variables. Thus, the indicators for S&T capacity are based upon our assessment of the factors that enable the absorption, retention, use, and creation of knowledge. We judged that including as many reliable indicators as we could find would increase the usefulness of the final index. Thus, we identified and collected together data on:

- Scientists and engineers trained at world class levels,
- Institutions where research is conducted,
- Public funds for research and development,
- Flows of information within the knowledge-using sectors ,
- Connectivity with the larger S&T world,
- The stock of embedded knowledge,
- Infrastructure to support economic and research activity.

All these data represent factors that are initial conditions that help to create S&T capacity. However, not all of the factors are equally contributory. It is possible to find an example of a country that has connectivity, infrastructure, and institutions where research could be conducted, but which does not have significant S&T capacity. However, it would be difficult to find an example of a country that has scientists and engineers, funds for R&D, and a stock of embedded knowledge that does not have the supporting features of institutions, information flows, and infrastructure.

Thus it could be argued that some features are *sufficient* to support S&T, while other features are *necessary* to support it. The necessary features are scientists and engineers, institutions for research, and funds for research and development. These three variables relate directly to S&T capacity. The other variables either measure the boundary conditions or environment of S&T or reflect the results of its application to scientific and technological production. We have selected eight quantitative indicators and have divided them into three distinct domains of S&T capacity, illustrated in Figure 2:

- **preconditions** that help create an environment conducive to the absorption, retention, production and diffusion of knowledge,

- **resources** of S&T activities, which concerns the indicators that relate most directly to S&T capacity,

- **output** of scientific and technological knowledge and its diffusion to the larger world.

The sources for the indicators are presented in Table 1.

## Coverage and comprehensiveness

Finding a balance between coverage (number of countries, regions or other units included in the analysis) and comprehensiveness (the variety of issues and dimension of science and technology) is perhaps the most challenging part of the task of creating an index. Obviously, no index can cover the entire range of dimensions related to a subject area. It is therefore advisable to sharply focus the index, that is, to define its precise purpose and to outline what it does and does not measure. Coverage is intimately related to comprehensiveness: the more detailed the data will have to be, the fewer countries, regions, or social groups can be included. This is especially true of many developing countries where fewer data are collected and statistical information is often less reliable.[2]

---

[2] The amount and reliability of such statistics cannot (necessarily) be blamed on the national statistical institutes. It is inherent to the nature of developing economies that many economic activities occur outside the statistical purview (e.g. payments in kind and production for own consumption).

A wide international comparison of a range of variables inevitably leads to problems regarding data availability. We sought to construct the index so we could include as many countries as possible, but the final list does not will not cover the entire world. Finding the balance between coverage and comprehensiveness entails three options:

- using fewer variables with a wider geographic coverage,
- reducing the sample of countries, keeping only those for which there is sufficient information,
- devising statistical methods to preserve both the entire sample of countries and the entire set of variables.

We have chosen the second option. This choice assumes that most S&T capacity is located in highly developed countries. As data becomes more and more scarce, the chances that we are missing S&T capacity also diminishes. As a result, we thought that we would lose more accuracy by eliminating variables than we could gain in breadth of coverage.

International statistical publications yielded a list of 215 countries, dependent and independent. We excluded 31 dependencies and small island nations for which data were extremely scarce. They include such countries as Tuvalu, Tonga and other Pacific island nations, Andorra and San Marino. The data coverage of the remaining 184 countries is shown in table 1.

Table 1

*Number of indicators covered by national data*

*and economic characteristics per group of countries*

| | | | | | | | | |
|---|---|---|---|---|---|---|---|---|
| number of indicators | 8 | 7 | 6 | 5 | 4 | 3 | 2 | 1 |
| number of countries | 66 | 17 | 37 | 22 | 4 | 5 | 32 | 1 |
| average per capita GDP ($) | 13,193 | 2,827 | 5,117 | 2,614 | 17,684 | 8,287 | 3,763 | a) |
| share in world population | 78.7 | 4.8 | 10.8 | 3.7 | 0.5 | 0.5 | 0.9 | a) |
| share in world GDP | 91.4 | 1.8 | 3.8 | 1.5 | 0.9 | 0.1 | 0.3 | a) |

a) There are no data on East Timor.

Is our sample of countries biased towards the high-income developed economies that can afford the collection of a large variety of statistical data? Have we adequately covered the developing countries?

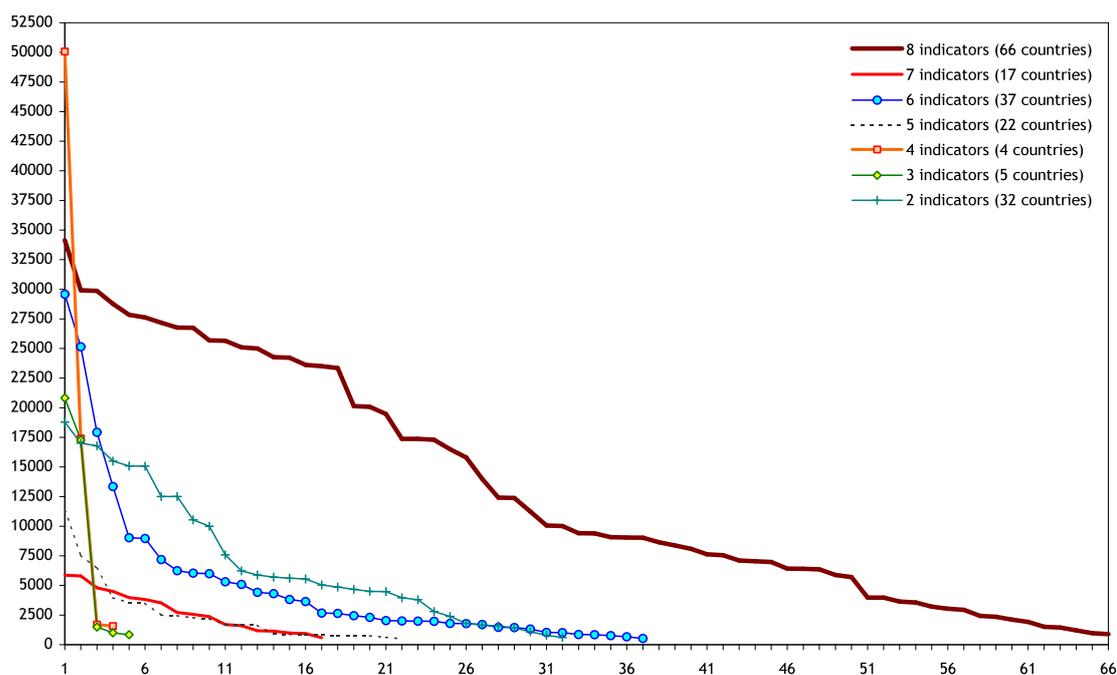

*Chart 3*

*Income distribution of countries*

*according to the number of indicators for which data are available*

**Conversion to a common format**

If the indicators are comparable —for example, when they are all values expressed in US dollars— then the composite index is simply equal to their sum or average. In most events the indicators will not be comparable and it will be necessary to first transform the percentages, values, ratios and other units to a common format. For example, the Human Development Index (HDI) of the UN Development Program converts its data on per capita GDP, life expectancy, and education to indices with a value between 0 and 1 before aggregating them into the HDI.

The variables included in the S&T Capacity Index are all different. Our preferred method of standardisation is to convert the absolute values into distances from an international average. The distance of the national score for a particular variable from the international average is expressed as a percentage of the standard deviation (the average of these distances). The formula for the conversion is:

$$Y_{ij} = \frac{X_{ij} - \overline{X}_j}{\sigma_j}$$

*Equation 1*

where:

$Y_{ij}$ is the converted value of indicator j for country i

$X_{ij}$ is the value of indicator j for country i

$\overline{X}_j$ is the international average of indicator j across the dataset

$\sigma_j$ is the standard deviation across the dataset of indicator j

The standard deviation and the international average serve as benchmarks in the comparison of the S&T capacity of countries. Since the ranking is entirely relative, in that the addition of a single country changes the scores of all countries and S&T capacity is not measured in absolute terms, it is very difficult to judge the Index value of a specific country. The international average and the margins defined by the standard deviation are used to classify the countries in the ranking as is shown in chart 4.

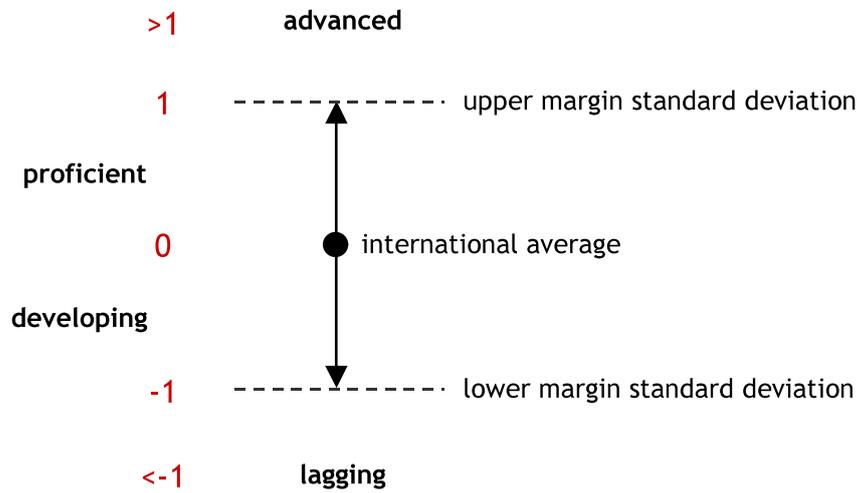

*Chart 4*

*A graphic representation of the interpretation of the S&T Capacity Index scores*

An alternative method of conversion that does result in an absolute score that is not dependent on the scores of other countries is the method of the Human Development Index. The basic formula of the HDI is:

$$HDI_{ij} = \frac{X_{ij} - X_{min}}{X_{max} - X_{min}}$$

*Equation 2*

where

$HDI_{ij}$ is the converted value of indicator j for country i

$X_{ij}$ is the value of indicator j for country i

$X_{min}$ is the lower boundary assigned to indicator j

$X_{max}$ is the upper boundary assigned to indicator j

The HDI method has its own shortcomings in that the upper and lower boundaries between which each indicator is assumed to move have a large impact on the ultimate value of the Capacity Index.

**Internal consistency**

One of the main requirements of a good composite index is that it has to be internally consistent. Each indicator has to have the same type of influence on the composite result. For example, if the value of one indicator can range from –1 to +1 while the value of another indicator ranges from 0 to +1, then the index is not internally consistent because the latter indicator can never have a negative influence. Another example of inconsistency is the combination of an indicator that makes a positive contribution to the composite index when its value *increases* with one that makes a positive contribution when it *declines*. Finally, different component variables can be substitutes or complementaries. For example, if two indicators are very closely connected you may actually be double-counting an effect (R close to 1) or inversely measuring it (R close to –1) in which case two indicators cancel each other out.

Three tests help determine the internal consistency of the index.

1. An analysis of the *distribution of values around the mean* shows the degree to which the aggregate is sensitive to variations in each variable. The distribution around the mean can be different for the component variables and it is important to know how this will affect the composite index.

2. The second test involves calculating the *correlations between the component variables* in order to discover substitutes and complementaries.

3. The third test relates to *consistency through time*. The volatility and growth rates of each component can vary considerably. If one component has a structurally higher average annual growth or stronger annual fluctuations, then over time and at any given moment its impact on the composite index can be more substantial and change more significantly than that of other indicators. Does the method of construction take this into account?

*Table #*

*Mean, median, standard deviation, and skewness*

*of the eight indicators in the S&T Capacity Index*

| | mean | median | standard deviation | skewness |
|---|---|---|---|---|

**Preconditions**

| | | | | |
|---|---|---|---|---|
| gross tertiary science enrolment ratio | 9.54 | 9.75 | 6.17 | .742 |
| per capita GDP | 13,193 | 9,409 | 9,648 | .470 |
| **Resources** | | | | |
| scientific engineers per million inhabitants | 1,461 | 1,320 | 1,286 | .718 |
| institutions per million inhabitants | 8.52 | 3.54 | 14.58 | 3.910 |
| R&D expenditure as a percentage of GDP | 1.04 | .73 | .87 | 1.045 |
| **Output** | | | | |
| Coauthorship Index | 437 | 167 | 652 | 2.550 |
| patents per million inhabitants | 31.16 | 1.34 | 56.36 | 2.215 |
| S&T journal articles | 218.18 | 92.70 | 273.19 | 1.269 |

As the indicators move from preconditions towards output the distribution around the mean becomes more skewed. Charts #a and #b show that enrolment and per capita GDP are spread fairly evenly across countries, although the normal distribution clearly tapers off towards the upper end of the scale. Resources are distributed somewhat more unevenly, which is especially true for the number of institutions. Output is particularly skewed as a large number of countries scores at or near zero. The distortion in the distribution of output indicators may be an artifact of the data. The three output indicators have a clear international bias and thus measure participation in global S&T creation and diffusion (part of the process of globalization) rather than national capacities.

The indicators have more or less the same range of values (table #). The original values were all positive and after conversion they range from about —2 to +6. The outliers are mainly found in the upper ranges and especially in the number of institutions, the coauthorship index, and the number of patents.

*Table #*

*Ranges of the converted values of the eight indicators*

|  | minimum | maximum |
|---|---|---|
| **Preconditions** | | |
| gross tertiary science enrolment ratio | -1.497 | 2.893 |
| per capita GDP | -1.280 | 2.273 |
| **Resources** | | |
| scientific engineers per million inhabitants | -1.132 | 2.825 |
| institutions per million inhabitants | -.576 | 6.495 |
| R&D expenditure as a percentage of GDP | -1.182 | 3.208 |

**Output**

| | | |
|---|---|---|
| Coauthorship Index | -.652 | 4.337 |
| patents per million inhabitants | -.554 | 4.025 |
| S&T journal articles | -.781 | 2.737 |

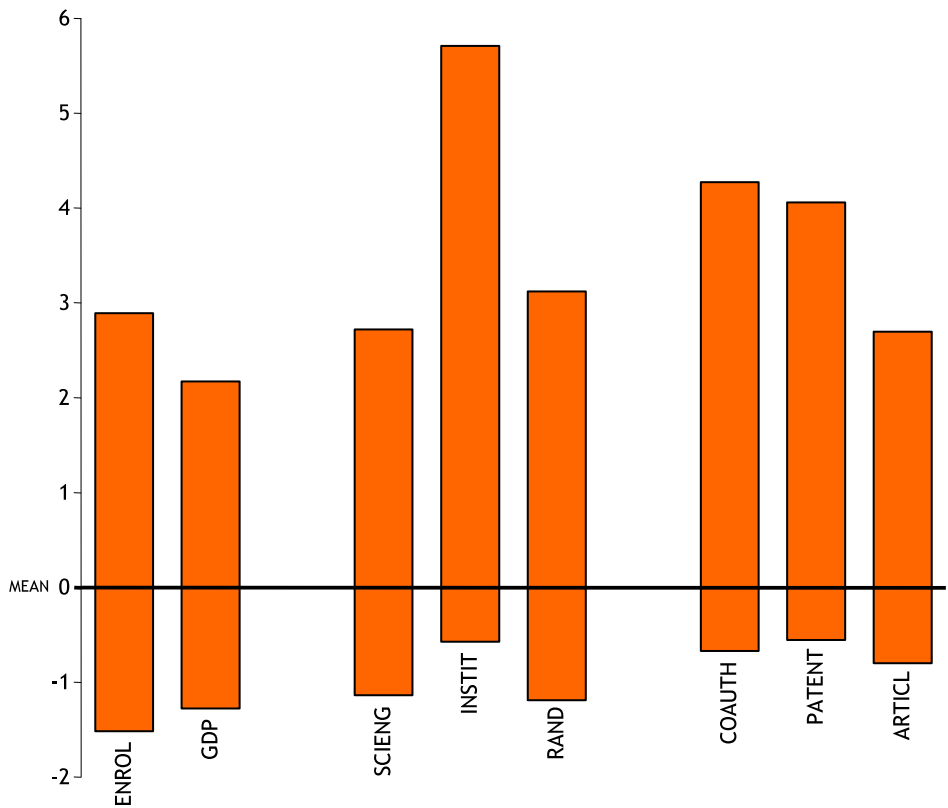

*Chart 5*

*Range of converted values around the mean*